\documentclass[a4paper]{article}
\usepackage{INTERSPEECH2020}
\usepackage{amsmath,graphicx, amssymb,tabularx}
\usepackage{multirow}
\usepackage{enumitem}
\usepackage{cite}

\newcolumntype{D}{>{\centering\arraybackslash}m{6ex}}

\title{Faster, Simpler and More Accurate Hybrid ASR Systems Using Wordpieces}
\name{
\begin{tabular}{c}
Frank Zhang, Yongqiang Wang, Xiaohui Zhang, Chunxi Liu, Yatharth Saraf, Geoffrey Zweig \thanks{The authors would like to thank Duc Le for helpful discussion about chenone, Jun Liu about CTC decoding and Abdelrahman Mohamed about transformer models}
\end{tabular}
}
\address{Facebook AI, USA}
\email{\{frankz,yqw,xiaohuizhang,chunxiliu,ysaraf,gzweig\}@fb.com}
\begin{document}
\ninept
\maketitle
\begin{abstract} 
In this work, we first show that on the widely used LibriSpeech benchmark, our transformer-based context-dependent connectionist temporal classification (CTC) system produces state-of-the-art results. We then show that using wordpieces as modeling units combined with CTC training, we can greatly simplify the engineering pipeline compared to conventional frame-based cross-entropy training by excluding all the GMM bootstrapping, decision tree building and force alignment steps, while still achieving very competitive word-error-rate. Additionally, using wordpieces as modeling units can significantly improve runtime efficiency since we can use larger stride without losing accuracy. We further confirm these findings on two internal \emph{VideoASR} datasets: German, which is similar to English as a fusional language, and Turkish, which is an agglutinative language.
\end{abstract}

\section{Introduction}
\label{sec:intro}

Deep neural networks have been the de facto architecture for automatic speech recognition (ASR) tasks since they were first introduced\cite{hinton2012deep}. These network architectures have evolved in recent years and can be broadly classified into two categories: 1) Those that support streaming during inference, such as time delay neural network (TDNN)\cite{peddinti2015time}, feed-forward sequential memory networks (FSMN)\cite{zhang2015feedforward}, long short-term memory (LSTM) \cite{sak2014long}, latency-controlled bi-directional LSTM (LC-BLSTM)\cite{zhang2016highway, le2019senones} and time-depth separable convolutions (TDS) \cite{hannun2019sequence} etc. and, 2) Full sequence architectures when latency is not a concern, e.g. BLSTM\cite{luscher2019rwth}, which can be used to provide better accuracy since the neural network can take full advantage of future information. Recently, Transformer\cite{vaswani2017attention} architectures have shown superior results in ASR tasks compared to BLSTMs\cite{wang2019transformerbased, karita2019comparative, sperber2018self,synnaeve2019endtoend}. In this work, we use LC-BLSTM as a representative for streaming and Transformer for the full sequence use case. 
    
Traditional hybrid DNN/Hidden Markov Model (HMM) approach utilizes a neural network to produce a posterior distribution over tied HMM states\cite{dahl2011large, mohamed2011acoustic} for each acoustic frame, usually followed by sequence discriminative training to boost performance\cite{povey2016purely}. CTC\cite{graves2006connectionist} has became an alternative criterion to frame-level cross-entropy (CE) training or sequence-level lattice-free MMI (LF-MMI) training in recent years and has shown promising results\cite{amodei2016deep,7404851,7178778,sak2015fast,45555}. Inspired by the rise of end-to-end training in machine translation, encoder-decoder architecture was also introduced for ASR, e.g. Listen, Attend and Spell (LAS) \cite{chan2016listen}. Most recently, neural transducers\cite{graves2012sequence} have shown great potential for both on-device\cite{sainath2020streaming} and server\cite{zhang2020transformer} use cases. Here, we consider both CE and CTC trained systems as hybrid since the neural network is solely modeling posteriors distribution over modeling units and a WFST-based decoder was used to produce hypothesis. While the LAS and Transducer based systems are considered end-to-end since there are decoder neural network components that directly produce hypotheses. In this work, our focus will be on CTC-based systems.

The most extensively studied modeling unit of hybrid ASR systems is tied context-dependent (CD) states/phones, i.e. senone\cite{dahl2011context}. As an alternative, chenone\cite{le2019senones} was proposed that has not only shown improvement in accuracy, but also eliminates the need of a phonemic lexicon. Since CTC training does not require alignment labels per frame, graphemes\cite{amodei2016deep}, wordpieces (WP)\cite{6289079,xiao2018hybrid,9003834,das2018advancing,synnaeve2019endtoend} or even whole words\cite{soltau2016neural} can be directly modeled.

In the rest of this paper, we first compare performance between chenone and wordpiece as modeling unit. We further study the best striding scheme for both modeling units and the impact of wordpiece vocab sizes. Then we perform neural language model rescoring on top of our best system to produce final WERs on LibriSpeech. With a transformer network trained with chenone-CTC, we achieve state-of-the-art result on LibriSpeech among hybrid systems. On our internal \emph{VideoASR} tasks,  a system trained with wordpiece-CTC only slightly lags behind in terms of WER, but is up-to 3x faster during inference compared to systems using chenone due to the larger stride in the neural network.

\section{Hybrid Architecture}
\label{sec:hybrid}
In hybrid ASR, an \emph{acoustic encoder} is used to encode a sequence of acoustic frames $\boldsymbol{x}_1, \cdots, \boldsymbol{x}_T$ to a corresponding sequence of high level embedding vectors $\boldsymbol{z}_1, \cdots, \boldsymbol{z}_T$. A softmax layer is then applied on these embedding vectors to produce a posterior distribution over the chosen modeling unit, e.g. senone or chenone, for each frame. These posterior distributions are then fed into a weighted finite-state transducer (WFST)\cite{mohri2002weighted} based decoder with a decoding graph also composed with a lexicon and language model (LM) to find the best hypothesis. 

In contrast to CE training that requires pre-computed per frame labels usually through a force alignment process, CTC training can implicitly learn the alignment between the input sequence and target sequence by introducing an additional blank label. The blank label is used to estimate the probability of outputting no label at a given frame. The encoder will be trained to produce probability distribution over all labels including blank. The log-likelihood of a given target sequence $\boldsymbol{y}$ can then be found by summing the probabilities of all allowed alignments. Specifically, 
\begin{align}
    \log p(\boldsymbol{y}|\boldsymbol{x}_1, \cdots, \boldsymbol{x}_T) = \sum_{\pi \in B^{-1}(\boldsymbol{y}) }\prod_{t=1}^{t=T}{p(\pi_t|\boldsymbol{x_t})}
\end{align}
where $B$ is the mapping operation that removes all blank and repeating labels in a given sequence. Note that the underlying assumption is probabilities between timestamps are conditional independent, which is ensured since the encoder is non-autoregressive. The network is then trained to maximize the log-likelihood for each training example and $p$ can be computed efficiently using forward-backward algorithm.

\section{Acoustic Model Architecture}
\label{sec:am}
In this section, we briefly review the neural network architectures we are going to study in this work: Transformer (in Section \ref{sec:transformer_arch}) and LC-BLSTM (in Section \ref{sec:lc_blstm_arch}).

\subsection{Architecture of Transformer}
\label{sec:transformer_arch}
Unlike when the Transformer\cite{vaswani2017attention} was first proposed as an encoder-decoder architecture for machine translation task, we only use the encoder part for acoustic modeling. Specifically we follow the setup in \cite{wang2019transformerbased} and use VGG layers\cite{abdel2014convolutional} in lieu of the original sinusoid positional encoding, since we have seen that for ASR tasks, convolutional positional encoding performs the best. Iterated loss\cite{Andros2019} was also applied to the intermediate embedding of transformer layers to help convergence and improve accuracy. Also different from the original transformer, we apply layer normalization \cite{lei2016layer} before multi-head attention (MHA) and feed-forward network (FFN) and we have an extra layer normalization operator after the residual connection. This is necessary to prevent bypassing the transformer layer entirely and helps with model convergence. 

\subsection{Architecture of LC-BLSTM}
\label{sec:lc_blstm_arch}

Unidirectional recurrent neural networks such as LSTMs base their predictions solely on the history they have already seen, hence the prediction accuracy is worse than bi-directional LSTMs that have access to the full context of the input, including future frames. However for streaming applications such as live captioning, we cannot wait for the full context to arrive because the ASR output needs to be made available within a certain latency budget of the audio stream being fed into the ASR system.

To strike a balance between recognition latency and accuracy, LC-BLSTM was first introduced in \cite{zhang2016highway}. Unlike BLSTM that cannot produce any hypothesis until the whole audio input is processed, LC-BLSTM only utilizes a limited number of right context ($RC$) frames to make predictions, which controls the latency. Similar to BLSTM, each LC-BLSTM layer also has two LSTMs, one left-to-right LSTM and one right-to-left LSTM. The difference is that the the input sequence is first divided into overlapping chunks of chunk size ($CS$) frames. The amount of overlap between chunks is equal to $RC$ frames. When forwarding left-to-right LSTM, hidden states and cell states are carried over between chunks, so that we have unlimited left context as in BLSTM. We forward right-to-left LSTM on each chunk and only keep $CS-RC$ frames of output activations, so that each input frame has at least $RC$ frames of right context to produce its activation. This mechanism enables LC-BLSTM to generate better acoustic embeddings than LSTM, without delaying the generation until the encoder has seen the whole audio.

\section{Modeling Units}

\subsection{Chenone}
\label{sec:chenone}

Chenone was first introduced in \cite{le2019senones} as an alternative to traditional senone \cite{dahl2011context} unit. Chenone not only eliminated the need for a phonetic lexicon, usually generated by linguists, but also showed WER improvement relative to phonetic modeling units in some cases. In order to use chenone as modeling unit in CTC training, we shift the labels after force alignment by one (all labels +1), and use label $\#0$ as the blank label in CTC. Essentially the neural network is now modeling distribution over all chenones plus one blank symbol. We then squeeze same adjacent labels in the alignment sequence into one label, let the encoder trained with CTC criterion to learn the implicit alignments. 

\begin{figure}[hhh]
    \centering
    \includegraphics[scale=0.15]{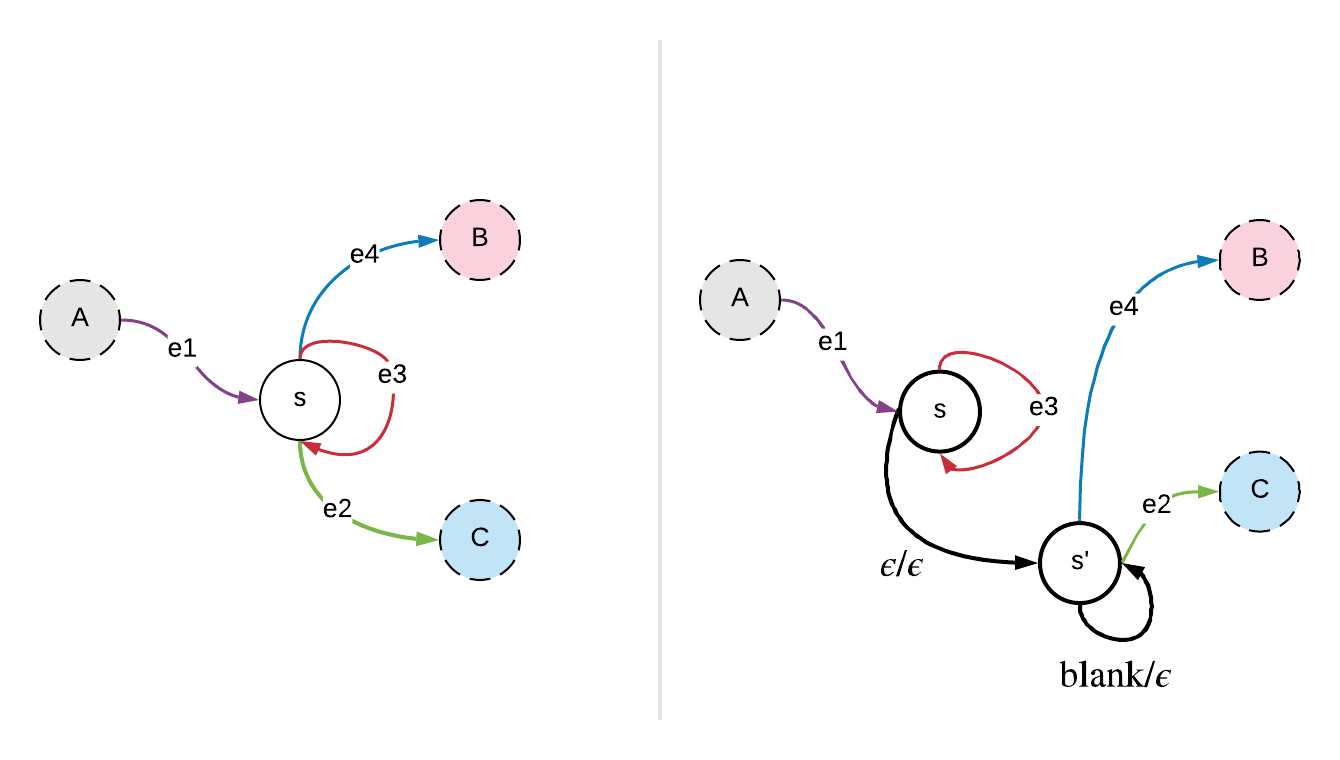}
    \caption{Convert a standard CD-HMM FST to be CTC compatible. A standard FST was shown on the left and a CTC compatible FST after conversion on the right}
    \label{fig:ctc_fst_convert}
\end{figure}

After a CTC model is trained, the next step is to construct a decoding graph. We first build an $\rm{H \circ C \circ L \circ G}$\footnote{Note that L here is simply a mapping between word and its graphemes. Examples could be found in \cite{le2019senones}}  graph following the standard procedure for CD-HMM (here, we always assume every HMM only has 1 state); then we transform it to a new graph which can consume an extra blank symbol. For each FST state $s$ in the decoding graph (shown in Figure \ref{fig:ctc_fst_convert}), we split that state into two FST states, $s$ and $s'$; we moved the outgoing edges ($e2$ and $e4$) to start from $s'$; A self loop edge with blank label as input and $\epsilon$ as output symbol with weight one (in the semi-ring's sense) is added; $s$ and $s'$ are then connected by a $\epsilon:\epsilon$ edge. Self-loop edges ($e3$ in this example) and incoming edge ($e1$) in this case are not changed. The transition model is on-the-fly converted to a new mapping function that shifts the output units by one. Once we change the transition model and decoding graph, the standard Kaldi decoder can be used to decode chenone-CTC model.

\vspace{-1em}
\vspace{-1.5em}

\subsection{Wordpiece}
\label{sec:wordpiece}

Subword unit representations such as byte pair encoding (BPE)\cite{sennrich2016neural} and wordpiece model\cite{WuSCLNMKCGMKSJL16} have been proposed with improved performance in many natural language processing (NLP) tasks. This approach chooses to divide words into a limited set of subword units, e.g. the word ``hello" may be encoded as ``\_he ll o", where the underscore indicates word start. In this work, we train wordpiece model using unigram language model word segmentation algorithm\cite{kudo2018subword}, and use the generated wordpiece vocab plus blank symbol as modeling units.

Since wordpiece is context independent, we only need to build an $\rm{H \circ L \circ G}$ graph for decoding, where: $H$ transduces $n+1$ symbols to $n$ wordpieces, i.e. absorbing the blank symbol; $L$ maps the sequence of wordpieces to the sequences of words, e.g. ``\_he ll o" to ``hello", and is done by the trained wordpiece model; $G$ is the standard $n$-gram word level LM. Once these FSTs are constructed, a standard procedure is used to compose $H$, $L$ and $G$ together. 

Compared to alignment-based training which first need to train a bootstrap model and build decision tree; then use the bootstrap model to perform force alignment and finally CE training, we only need to train a wordpiece model using text-only data and followed by one-stage CTC training. The whole training pipeline is greatly simplified.

\section{Experiments}
\label{sec:exp}
To evaluate the performance of different modeling units and training criterion, we first conduct experiments on the LibriSpeech corpus\cite{panayotov2015librispeech}, followed by experiments on two larger and more challenging internal datasets of German and Turkish social media videos.

\subsection{Data}
\label{sec:dataset}
The LibriSpeech corpus contains about 960 hours of read speech data for training, and 4 development and test sets (\emph{\{dev, test\}-\{clean,other\}}), where \emph{other} sets are more acoustic challenging. We use the official 4-gram language model (LM) with 200K vocabulary for all first-pass decoding and n-best generation for neural LM rescoring. 

The German and Turkish datasets are our in-house video datasets, which are sampled from public social media videos. The datasets are completely de-identified before transcription; both transcribers and researchers do not have access to any user-identifiable information (UII). The training, validation and test set sizes are shown in Table \ref{tab:video_data}. All hyper-parameter tuning is done on validation set. The video datasets contain a diverse array of speakers, accents, video categories, and acoustic conditions, and are more challenging than the LibriSpeech dataset explored in this work.


We use the same bootstrap model obtained from decision tree building stage to segment the training split of both LibriSpeech and video dataset to up to 10 seconds \footnote{
This is achieved by force aligning the whole audio against the reference using the LC-BLSTM acoustic model. }. We also segment the dev and test split of video dataset to up to 10s, while no segmentation is performed on LibriSpeech dev and test sets in order to be comparable with other published results. The benefit of segmenting training data was shown in \cite{wang2019transformerbased}.

\begin{table}[htb]
    \centering
    \caption{Dataset sizes for internal Video ASR tasks. Number of utterances in parentheses.}
    \begin{tabular}{*3c}
    \hline
    Language & German & Turkish \\
    \hline\hline
    train & 3K hrs (\textasciitilde 135K) & 3.1K hrs (\textasciitilde 137K) \\
    valid & 14.5 hrs (\textasciitilde 600) & 14.4 hrs (\textasciitilde 600)  \\
    test  & 24.2 hrs (\textasciitilde 1K) & 24.4 hrs (\textasciitilde 1K)  \\
    \hline
    \end{tabular}
    \label{tab:video_data}
\end{table}
\vspace{-1em}

\subsection{Experiment Setup}

We follow \cite{le2019senones,wang2019transformerbased} to use context- and position-dependent graphemes (i.e., \textit{chenones}) for CE baselines and CTC experiments. We bootstrap our HMM-GMM system using the standard Kaldi \cite{Povey_ASRU2011} LibriSpeech recipe. We use 1-state HMM topology with fixed self-loop and forward transition probability (both 0.5). 80-dimensional log Mel-filter bank features are extracted with a 10ms frame shift and 25ms FFT windows. Speed perturbation and \emph{SpecAugment} \cite{park2019specaugment} (LD policy without time warping) are applied to all experiments unless specially noted.

Since the focus of this work is not on network architecture searches, we use our previously found best setup across all following experiments. For acoustic transformers, we use a 24-layer transformer encoder architecture with embedding dimension 512 and 8 attention heads; the FFN dimension is 2048. Iterative loss was applied to intermediate output embeddings of layer 6, 12 and 18. We use three VGG blocks \cite{simonyan2014very} to encode acoustic features before feeding into transformer layers: each VGG block contains 2 consecutive convolution layers with a 3-by-3 kernel followed by a ReLu non-linearity and a pooling layer; 64 channels are used in the convolution layer of the first VGG block and increase to 128 for the second block and 256 for the third block. Max-pooling is performed at a 2-by-2 grid, with optional stride choice from 1 to 3 in each block. This model has about 81M parameters and we note the model as \emph{vggTrf}. For LC-BLSTMs, we follow \cite{le2019senones} and use 5 layers with 800 hidden units per layer per direction. Optional subsampling by a factor of 2 or 3 can be applied after the first hidden layer. This model has about 83M parameters, similar to the transformer model. Dropout is applied in all experiments: 0.1 for transformers and 0.2 for LC-BLSTM. 

All neural network training is performed using an in-house developed \emph{PySpeech} framework that is built on top of the open-sourced PyTorch-based \emph{fairseq}\cite{ott2019fairseq} toolkit. Adam optimizer \cite{kingma2014adam} with (0.9, 0.999) betas and $1e^{-8}$ epsilon is used in all experiments. We apply the \emph{tri-stage}\cite{park2019specaugment} learning rate (LR) scheduler. The hold stage LR is $1e^{-3}$. For experiments on transformer models, we follow a schedule of (48K, 100K, 200K) steps with a batch contains up to 20,000 frames. For LC-BLSTM models we use schedule of (16K, 32K, 64K) steps with a batch contains up to 50,000 frames. All models are trained on 32 Nvidia V100 GPUs for 200 epochs in total. Training is usually finished between 2 to 4 days. 

Test set WERs are obtained using the best model based on evaluated WER on the development set. The best checkpoints for both LibriSpeech \emph{test-clean} and \emph{test-other} are selected using the \emph{dev-other} development set. For video, the best checkpoint was selected using the valid set.

\subsection{Effect of Stride}
\label{ssec:stride}
In the first set of experiments, we investigate what is the best stride for different modeling units. For \emph{vggTrf} model, striding is achieved by setting stride of max-pooling layers, e.g. a total stride of 8 can be achieved by setting stride equals 2 for all three VGG blocks. For LC-BLSTM, we apply subsampling factor of 2 for activation of the three consecutive hidden layers right after the first layer. We follow the setup in \cite{le2019senones} and use stride 2 for the chenone-CE baseline. We use 2K wordpiece size in this study. Results on LibriSpeech \emph{test-other} dataset shown in Table \ref{tab:stirdes}.

\begin{table}[htb]
    \centering
    \caption{Effect of stride for wordpiece and chenone}
    \begin{tabular}{|c|c|c|cc|}
    \hline
    Unit & Criterion & Stride & \emph{LC-BLSTM} & \emph{vggTrf} \\
    \hline\hline
    chenone & CE & 2 & 8.81 & 5.59 \\
    \hline
    \multirow{5}{*}{WP} & \multirow{5}{*}{CTC} & 2 & 9.75 & 5.96 \\
     & & 2*2 & \textbf{8.92} & \textbf{5.74} \\
     & & 3*2 & 9.16 & 5.84 \\
     & & 2*2*2 & 8.94 & 5.90 \\
     & & 3*2*2 & 9.34 & 6.51 \\
    \hline
    \multirow{4}{*}{chenone} & \multirow{4}{*}{CTC} & 2 & 9.43 & 5.28 \\
     & & 2*2 & \textbf{8.51} & \textbf{5.16} \\
     & & 3*2 & 9.93 & 6.35 \\
     & & 2*2*2 & 17.94 & 17.27 \\     
    \hline
    \end{tabular}
    \label{tab:stirdes}
\end{table}

The results show that for chenone-CTC, the best stride is 4 and it out performs CE baseline. While for WP-CTC system, we find that we can use stride as high as 8 without losing much accuracy. Because wordpiece model is trained based on word frequency, some common whole words, e.g. 'welcome' and 'information', made into the vocab. The acoustic model can use larger stride since the corresponding input frames will span longer. On the other hand, although context dependent, chenones are essentially still characters and have a shorter corresponding time span in input sequence. So we observed that WER degrades quickly when larger stride is used.

\subsection{Effect of Wordpiece Dictionary Size} 

In the second set of experiments, we explore the effect of different wordpiece sizes. Here, all models use a total stride of 8. We report results on LibriSpeech \emph{test-other} dataset in Table \ref{tab:wp_sizes}.

\vspace{-0.5em}
\begin{table}[htb]
    \centering
    \caption{WER using different wordpiece sizes}
    \begin{tabular}{|c|cc|}
    \hline
    Wordpiece size & \emph{LC-BLSTM} & \emph{vggTrf} \\
    \hline\hline
    1K & \textbf{8.71} & \textbf{5.86} \\
    2K & 8.94 & 5.90 \\
    5K & 9.13 & 6.05 \\
    10K & 9.23 & 6.35 \\
    16K & 9.44 & 6.51 \\
    \hline
    \end{tabular}
    \label{tab:wp_sizes}
\end{table}
\vspace{-0.5em}

The results show that on LibriSpeech, smaller wordpiece vocab size tends to work better. It's worth exploring vocab sizes below 1K in future study. We also explored subword regularization during training following \cite{kudo2018subword}. We tried the setup of $(l=64, \alpha=0.1/0.5)$ but results turn out slightly worse. During decoding we also tried building decoding graph with multiple pronunciations in the lexicon with corresponding pronunciation probability for the lexicon entries. The results turn out on-par. We also explored sMBR and mWER training after CTC stage. It provides very marginal gain so we didn't include the results in this study. 

To achieve best WERs, we trained a 36-layer stride-4 \emph{vggTrf} model with chenone-CTC(about 124M parameters), and changed time masking of \emph{SpecAugment} LD policy to $(T=30, mT=10)$. Its performance and those of some other published LibriSpeech systems can be found in Table \ref{tab:comp}. The new system outperforms our previous best hybrid system\cite{wang2019transformerbased} by 11\% and 14\% respectively on \emph{test-clean} and \emph{test-other}. We also trained a 42-layer transformer LM following the setup in\cite{Irie_2019} with the LibriSpeech transcriptions and 800M-word text-only data. The transformer LM achieved perplexity 52.35 on the dev set (a combination of \emph{dev-clean} and \emph{dev-other}). We then perform n-best rescoring on up to 100-best hypotheses generated by first pass decoding. The oracle error rate of the n-best hypotheses are 1.0\% and 2.2\% on \emph{test-clean} and \emph{test-other} respectively. Our final WERs (2.10\%/4.20\%) are the best results among all hybrid systems on this widely used benchmark.

\begin{table}[htb]
    \centering
    \caption{Comparison of our chenone-CTC with previous best results on LibriSpeech. ``4g" means the official 4-gram LM was used; ``NNLM" means a neural LM was used.
    }
    \begin{tabular}{|c|c|c|DD|}
    \hline
    Arch. & System & LM & test-clean & test-other \\
    \hline\hline
     \multirow{3}{*}{LAS} & Karita et al.\cite{karita2019comparative} & NNLM & 2.6 & 5.7 \\
        & Park et al.\cite{park2019specaugmentv2} & NNLM & 2.2 & 5.2 \\
        & Synnaeve et al.\cite{synnaeve2019endtoend} & NNLM & 2.33 & 5.17 \\        
    \hline\hline
     \multirow{2}{*}{Transducer} & \multirow{2}{*}{Zhang et al.\cite{zhang2020transformer}} & No LM & 2.4 & 5.6 \\
        &  & NNLM & \textbf{2.0} & 4.6 \\
    \hline\hline
     \multirow{8}{*}{Hybrid} & \multirow{2}{*}{RWTH\cite{luscher2019rwth}} & 4g & 3.8 & 8.8  \\
        &  & +NNLM & 2.3 & 5.0 \\
        \cline{2-5}
        & \multirow{2}{*}{Han et al.\cite{han2019state}} & 4g & 2.9 & 8.3 \\
        &  & +NNLM & 2.2 & 5.8 \\
        \cline{2-5}
        & \multirow{2}{*}{Wang et al.\cite{wang2019transformerbased}} & 4g & 2.60 & 5.59 \\
        &                       & +NNLM & 2.26 & 4.85 \\
        \cline{2-5}
        & \multirow{2}{*}{Ours} & 4g & \textbf{2.31} & \textbf{4.79} \\
        &                       & +NNLM & 2.10 & \textbf{4.20} \\
    \hline
    \end{tabular}
    
    \label{tab:comp}
\end{table}
\vspace{-1.0em}

\subsection{Experiments on Video dataset}
\vspace{-0.5em}
Finally, we tested \emph{vggTrf} with CTC criterion on the more challenging and larger scale internal \emph{VideoASR} tasks, as described in Section \ref{sec:dataset}. Stride 2 is used for CE, 4 for chenone-CTC and 8 for WP-CTC training. Results are shown in Table \ref{tab:video_asr}. We find that both CTC systems outperform the baseline CE system. It's also consistent that chenone-CTC outperforms WP-CTC in terms of WER. On the other hand, WP-CTC system is significantly better in terms of real-time factor (RTF) due to the larger stride used. Blank frame skipping during decoding also contributes to the speedup. We skip frames if blank label posterior is greater than 99\%. Empirically, we found that over 20\% of the frames in chenone-CTC and over 50\% of frames in WP-CTC were skipped.
\vspace{1em}

\vspace{-1em}
\begin{table}[htb]
    \centering
    \caption{Experiment results on internal VideoASR tasks}
    \begin{tabular}{*6c}
    \hline
    \multirow{2}{*}{Unit} & \multirow{2}{*}{Criterion} & \multicolumn{2}{c}{German}  & \multicolumn{2}{c}{Turkish} \\
     &  & WER & RTF & WER & RTF \\
    \hline\hline
    chenone & CE & 15.54 & 0.26 & 21.92 & 0.25 \\
    WP & \multirow{2}{*}{CTC} & 14.32 & \textbf{0.07} & 19.04 & \textbf{0.06}   \\
    chenone & & \textbf{13.74} & 0.10 & \textbf{18.17} & 0.10  \\
    \hline
    \end{tabular}
    \label{tab:video_asr}
\end{table}
\vspace{-1em}


\section{Discussions and Conclusions}
\label{sec:con}

In this work, we pushed the performance boundary of hybrid ASR using transformer-based acoustic models with context-dependent CTC training. Modeling choices are discussed and compared in detail, and as Table \ref{tab:comp} shows, our system yields state-of-the-art results on the LibriSpeech benchmark.

For real world production system however, we must take into account not only recognition accuracy, but also engineering complexity, inference efficiency, flexibility etc. We have shown that hybrid systems can be built with fewer steps leveraging wordpiece and CTC training. Yet, the system's accuracy is very competitive with a much faster runtime inference speed. Results on a more challenging internal dataset show similar results, confirming that such a wordpiece-CTC system indeed has great potential.

\footnotesize
\bibliographystyle{IEEEtran}
\bibliography{strings,refs}

\end{document}